\documentclass{article}

\usepackage[pdftex]{graphicx}
\graphicspath{{./graphics}}
\DeclareGraphicsExtensions{.pdf}

\usepackage{amsmath}
\usepackage{amssymb}
\usepackage{algorithmic}
\usepackage{array}
\usepackage{color}
\usepackage[hidelinks]{hyperref}
\usepackage[tableposition=top]{caption}
\usepackage{subcaption}
\usepackage{cite}
\usepackage[table,xcdraw]{xcolor}
\usepackage{spconf}
\usepackage{multirow}
\usepackage[symbol]{footmisc}

\newcommand{\compresslist}{%
    \setlength{\topsep}{\parskip}
    \setlength{\itemsep}{1pt}
    \setlength{\parskip}{0pt}
    \setlength{\parsep}{0pt}
}

\usepackage{enumitem}
\setlist[itemize]{leftmargin=*}
\setlist[enumerate]{leftmargin=*}

\newlength{\bibitemsep}
\newlength{\bibparskip}
\let\oldthebibliography\thebibliography
\renewcommand\thebibliography[1]{%
  \oldthebibliography{#1}%
  \setlength{\parskip}{\bibitemsep}%
  \setlength{\itemsep}{\bibparskip}%
}

\begin{document}


\title{An Empirical Study of Conv-TasNet}

\name{Berkan~Kad\i o\u{g}lu$^{\hspace{1mm}  \ddagger \hspace{1mm} \star}$ \quad Michael~Horgan$^{\dagger \hspace{1mm} \star }$ \quad Xiaoyu~Liu$^{\dagger \hspace{1mm} \star }$ \quad Jordi~Pons$^{\dagger}$ \quad Dan~Darcy$^{\dagger}$ \quad Vivek~Kumar$^{\dagger}\thanks{$^\star$ \hspace{0.31mm} Work done during Berkan~Kad\i o\u{g}lu's internship at {Dolby Laboratories.} Berkan, Michael and Xiaoyu equally contributed to this work.}$}

\address{$^{\ddagger}$ Electrical and Computer Engineering Department, Northeastern University \\
$^{\dagger}$ Dolby Laboratories}

\markboth{Submitted to ICASSP 2020}%
{Shell \MakeLowercase{\textit{et al.}}: Title}%
\maketitle
\ninept

\begin{abstract}
Conv-TasNet is a recently proposed waveform-based deep neural network that achieves state-of-the-art performance in speech source separation. Its architecture consists of a learnable encoder/decoder and a separator that operates on top of this learned space. Various improvements have been proposed to Conv-TasNet. However, they mostly focus on the separator, leaving its encoder/decoder as a (shallow) linear operator. In this paper, we conduct an empirical study of Conv-TasNet and propose an enhancement to the encoder/decoder that is based on a (deep) non-linear variant of it. In addition, we experiment with the larger and more diverse LibriTTS dataset and investigate the generalization capabilities of the studied models when trained on a much larger dataset. We propose cross-dataset evaluation that includes assessing separations from the WSJ0-2mix, LibriTTS and VCTK databases. Our results show that enhancements to the encoder/decoder can improve average SI-SNR performance by more than 1 dB. Furthermore, we offer insights into the generalization capabilities of Conv-TasNet and the potential value of improvements to the encoder/decoder.

\end{abstract}
\section{Introduction}
\label{sec:introduction}

With the recent advent of deep learning, speech separation methods have experienced steadfast success in difficult scenarios where, e.g., prior information about the speakers is not available. Depending on the models' input/output, one can roughly categorize these methods into spectrogram- and waveform-based models. Spectrogram-based models, despite their success in the past 
\cite{kolbaek2017multitalker, hershey2016deep, liu2019divide}, have limitations: (i)~they discard phase information via simply estimating masks that operate over the magnitude or power spectrogram; (ii) they tend to employ the noisy phase of the mixture for reconstructing the clean source; and (iii) they employ a generic transform (like STFT) which might not be optimal for the task at hand. Although several works investigate how to address the above-mentioned limitations \cite{griffin1984signal, le2019phasebook, tan2019complex, rethage2018wavenet}, recent publications have reported promising results by tackling source separation directly in the waveform domain \cite{luo2019conv, slizovskaia2019end, shi2019deep, shi2019endfurca, stoller2018wave, lluis2018end, grais2018raw}. 

The Conv-TasNet \cite{luo2019conv} architecture, the work on which this paper builds, is one such end-to-end neural network that achieves state-of-the-art performance in speech source separation. Its architecture consists of two parts: an encoder/decoder, and a separator. %
Recently, several improvements have been proposed to this architecture.
However, most focus has been devoted to its separator. For example, in \cite{shi2019deep, shi2019endfurca} a parallel and multi-scale separator is proposed, and in \cite{yang2019improved} a clustering mechanism is integrated into the separator. Interestingly, only a few works touch on the encoder/decoder of Conv-TasNet. In a multi-channel setting \cite{gu2019end, bahmaninezhad2019comprehensive} a second encoder is used to learn phase differences between channels, and in \cite{yang2019improved} a magnitude STFT is appended to the learned encoder transform. As seen, most previous works use a (shallow) linear encoder/decoder. To the best of our knowledge, only \cite{pandey2019tcnn} used a deep encoder/decoder for a Conv-TasNet inspired model for speech enhancement, which has not been extended to or fully tested for speech source separation.

In this work, we conduct an empirical study of Conv-TasNet, which is formally introduced in Section 2. Our contributions focus on two areas: architectural improvements to the encoder/decoder, and a study of the generalization capabilities of the developed models. 
In Section 3, we introduce the deep encoder/decoder we propose and we discuss several variants of this structure. In Section 4.1, we evaluate the studied models against the WSJ0-2mix database to gain insights on the  performance of each variant. In Section 4.2,  we explore the impact of using a larger, more diverse training set and we study the generalization capabilities of the trained models via employing a cross-dataset evaluation.  In Section 4.3, we compare the performance of the proposed deep encoder/decoder to several state-of-the-art separators.  We conclude our discussion in Section 5.

\section{Review of Conv-TasNet}
\label{sec:review_of_conv-tasnet}

\begin{figure}[h]
    \centering
    \hbox{\hspace{1.5mm}\includegraphics[scale=0.27]{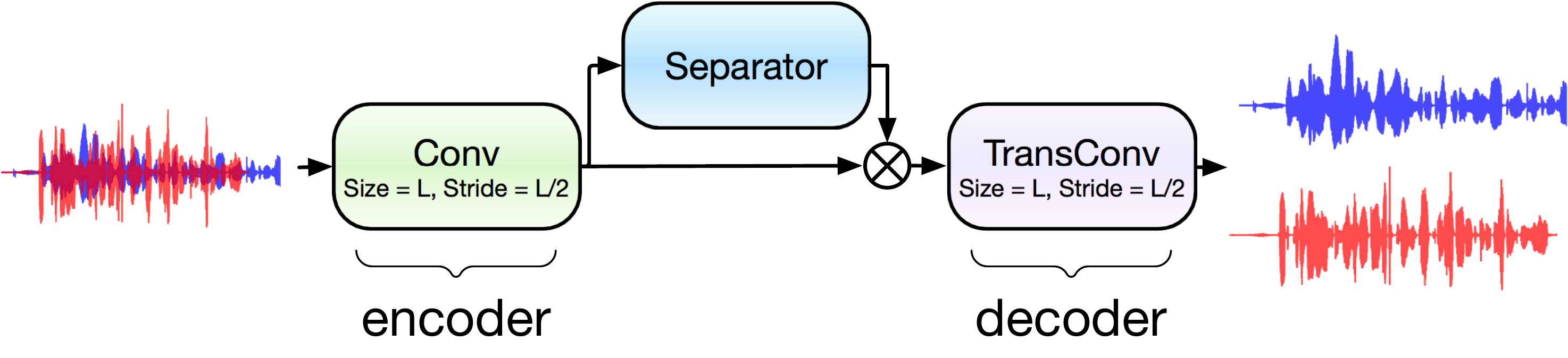}}
            \vspace{-3mm}
    \caption{The building blocks of Conv-TasNet.}
    \label{fig:ctnblocks}
        \vspace{-2mm}
\end{figure}

Single-channel multi-speaker speech separation aims to separate $C$ individual speech sources $\mathbf s_c \in \mathbb{R}^T$, where
\mbox{$~c~\in~\{1, 2,~\dots~,~C\}$}, from a single-channel mixture of speech \mbox{$\mathbf x \in \mathbb{R}^T$} where $T$ is the length of the waveform and $\mathbf x = \sum_{c=1}^C \mathbf s_c$. 
Conv-TasNet~\cite{luo2019conv} is an end-to-end fully convolutional network proposed for this purpose. Fig. \ref{fig:ctnblocks} illustrates the two main modules in Conv-TasNet: an encode/decoder pair, and a separator.

\begin{figure*}[!t]
    \centering
    \vspace{-5mm}
    \begin{subfigure}[]{\textwidth}
        \centering
        \includegraphics[scale=0.29]{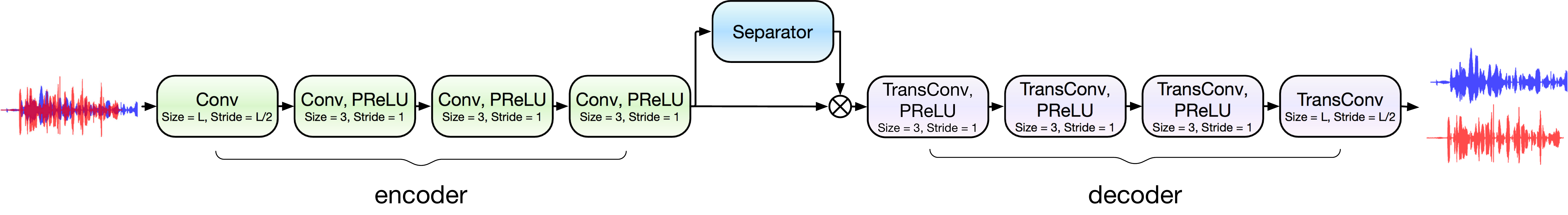}
    \end{subfigure}%
        \vspace{0mm}

    \caption{The Deep Encoder/Decoder Architecture.}
    \label{basicdeep}
        \vspace{-2mm}

\end{figure*}

\noindent The encoder linearly maps a mixture waveform into a learned latent space.  Specifically, a mixture waveform is segmented into $K$ overlapping frames $\mathbf x_k \in \mathbb{R}^L$, $k = 1, 2, \dots, K$, each of length $L$ with stride $S$.  Then, the linear transform is defined as:
\begin{equation} \label{linearenc}
   \mathbf E = \mathbf U\mathbf X,
\end{equation}
where $\mathbf X \in \mathbb{R}^{L \times K}$ stores all frames in columns, $\mathbf U \in \mathbb{R}^{N \times L}$ contains $N$ learnable basis filters in its rows, and $\mathbf E \in \mathbb{R}^{N \times K}$ is the latent space representation of the mixture waveform.  In practice, this encoder is implemented as a 1-D convolution with $N$ kernels.  In this work, $N=512$, $L=16$, and $S=8$, corresponding to 2 ms basis filters and 1 ms stride at a sample rate of 8 kHz.

The separator predicts a representation for each source by learning a mask in this latent space.  The temporal convolutional network (TCN) \cite{lea2016temporal} architecture is the core of the separator.  In Conv-TasNet, the TCN consists of 8 stacked dilated convolutional blocks with exponentially increasing dilation factors, and each stack is repeated 3 times.  A 
deep stack of dilated convolutions
enables the separator to have a large temporal context with a compact model size.

The decoder linearly transforms the latent space representation of each estimated clean source $c = 1,2,\dots,C$ to the time domain:
\begin{equation} \label{lineardec}
   \mathbf{\hat S}_c = \mathbf D^T_c\mathbf V,
\end{equation}
where $\mathbf V \in \mathbb{R}^{N \times L}$ contains $N$ decoder basis filters (not tied with the encoder $\mathbf U$), $\mathbf D_c \in \mathbb{R}^{N \times K}$ is the representation of the $c$th estimated source predicted by the separator, and $\mathbf{\hat S_c} \in \mathbb{R}^{K \times L}$ contains $K$ frames of the reconstructed signal.  The entire time domain waveform $\mathbf{\hat s}_c$ is obtained by overlap-and-add of the rows of $\mathbf{\hat S}_c$.  Similar to the encoder, the decoder is implemented as a 1-D (transposed) convolution.  All the Conv-TasNet building blocks are jointly optimized.
\section{Deep Encoder / Decoder}
\label{sec:deepencdec}

In this section, we describe the deep encoder/decoder architectures we used to explore the impact of increasing the Conv-TasNet encoder/decoder's capacity to represent more complex signal transformations.
The core architecture we employ is motivated by recent research in audio classification in which waveform-based architectures built on a deep stack of small filters deliver very competitive results~\cite{pons2019randomly,pons2018end,lee2017sample}.  This research highlights the potential for these architectures to learn generalized patterns via hierarchically combining small-context representations~\cite{pons2019randomly}.
For this reason, we investigate the possibilities of a deep encoder/decoder that is based on a stack of small filters with nonlinear activation functions.

\subsection{Deep Encoder/Decoder Architecture}
Fig. \ref{basicdeep} depicts the diagram of the proposed deep encoder/decoder. We utilize a nonlinear deep encoder with $I$ layers. The first layer is equivalent to the original Conv-TasNet encoder in Eq.1: a linear transformation is applied to frames of length $L$ and stride $S$. It is implemented via a 1-D convolutional layer with $N$ kernels.
The second part consists of a stack of~$I-1$ 1-D convolutional layers, with each layer having $N$ kernels of size 3 and a PReLU:
\begin{equation}
    \mathbf E_i = PReLU(\mathbf U_i * \mathbf E_{i-1}),
\end{equation}
where $*$ denotes the convolution operator, $i=2,3,\dots,I$ denotes the layer index, $\mathbf U_i \in \mathbb{R}^{N \times N \times 3}$ are the kernels, and $\mathbf E_i \in \mathbb{R}^{N \times K}$ is the layer output.  This deep stack of encoding layers hierarchically transforms the waveform into a nonlinear latent space.

The deep decoder directly mirrors the encoder architecture.  Masked encodings from the separator are first processed by $I-1$ dimension-preserving 1-D transposed convolutional layers with PReLU activations.  Finally, the linear kernel filters are applied via a transposed convolution as in Eq. \ref{lineardec}, with kernel size $L$ and stride $S$, to produce the time-domain estimated source signals.

\subsection{Deep Encoder/Decoder Variants}

In addition to the standard deep encoder/decoder architecture described above, we also examine two variants.  The first uses dilations in the deep convolutional layers of the encoder/decoder to increase its temporal receptive field.  As \cite{luo2019conv} points out, dilation is crucial for the separator to model long contexts. For this reason, we also examine its effect in the encoder/decoder.  In this variant, we experiment with up to 4 dilated nonlinear layers ($I=5$) with an exponentially increasing dilation factor (as in the separator), i.e.: $1, 2, 4, 8$ for the encoder and $8, 4, 2, 1$ for the decoder.

\begin{figure}[h]
    \centering
    \includegraphics[scale=0.3]{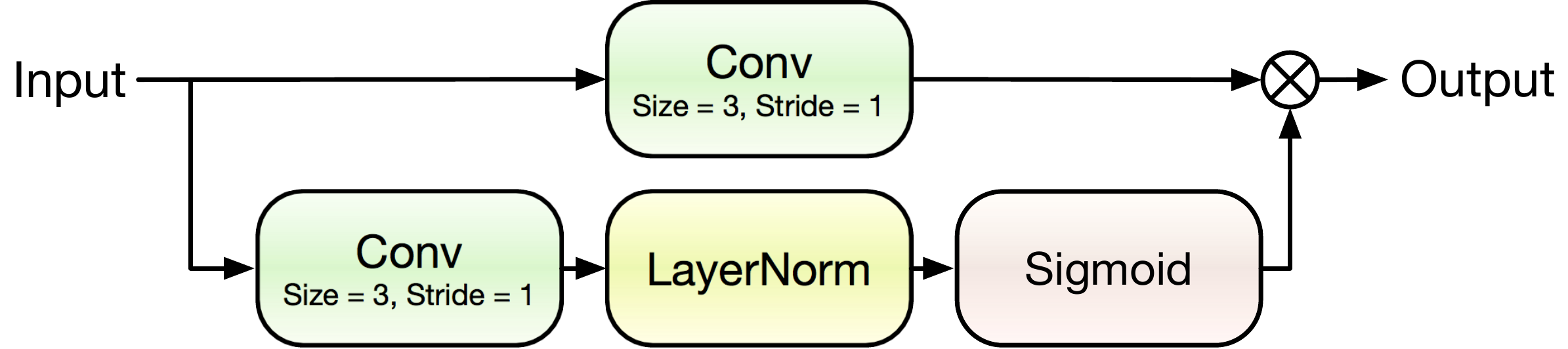} 
    \caption{The Gated Linear Unit used in this work.}
    \label{fig:GLU}
    \vspace{-5mm}
\end{figure}

The second variant uses gated linear units (\text{GLUs}) to replace the PReLU in the basic deep encoder/decoder.  Similar to attention, GLUs (Fig. \ref{fig:GLU}) rely on a learned gate to model the relative importance of the kernels.  A global layer normalization \cite{luo2019conv} is inserted before the sigmoid nonlinearity in the GLU to speed up training. Gated units have been shown to be effective for audio event detection \cite{xu2018weakly}, audio generation \cite{oord2016wavenet}, and speech enhancement \cite{tan2018gated}.

\subsection{Objective Functions}
By default, we use the scale-invariant signal-to-noise ratio (\text{SI-SNR}) \cite{le2019sdr} with permutation-invariant training \cite{kolbaek2017multitalker} as our objective function.  SI-SNR is a widely used objective function for end-to-end speech source separation \cite{luo2019conv,shi2019deep,shi2019endfurca,yang2019improved}.  
In certain experiments (see Section \ref{cdseval}), in an effort to constrain the scale of the predicted sources from the deep encoder/decoder, we augment the objective function with a power-law term that encourages the model to predict spectra that are of similar magnitude to the ground truth. Besides, power laws are well known to correlate with human perception \cite{stevenspowerlaw,kim2019feature}.  The augmented objective we use is:
\begin{equation} \label{plaw}
    \hspace{-4mm} L = -\text{SI-SNR}(\hat s_{c}, s_{c}) + \beta \cdot \text{P-law}(\hat s_{c}, s_{c}, \alpha),
\end{equation}
where
\begin{equation}
 \text{P-law}(x, y, \alpha) = \text{L1-norm}(||\text{STFT}(x)||^\alpha - ||\text{STFT}(y)||^\alpha).
\end{equation}

For our experiments with the P-law augmented objective we set $\beta$ to 0.01 and $\alpha$ to 0.5.

\section{Experiments}
\label{sec:experiments}
In this section we evaluate the deep encoder/decoder architecture and its variants, and we assess the impact of using different architectures, training sets, and objective functions on generalization.

For all of our experiments, we train on 4-second mixture utterances --- where a mixture consists of two clean speech sources from different speakers.  As in \cite{luo2019conv}, we use the ADAM optimizer with a learning rate of 1e-3, and a schedule that halves the learning rate after 3 consecutive epochs with no reduction in validation loss. We set our batch size to 16. We clip gradient norms to 7. Unless stated otherwise, we train for 100 epochs on WSJ0-2mix and 60 epochs on LibriTTS, and the objective function is SI-SNR.

\begin{figure}[!t]
    \centering
    \includegraphics[width=0.5\textwidth]{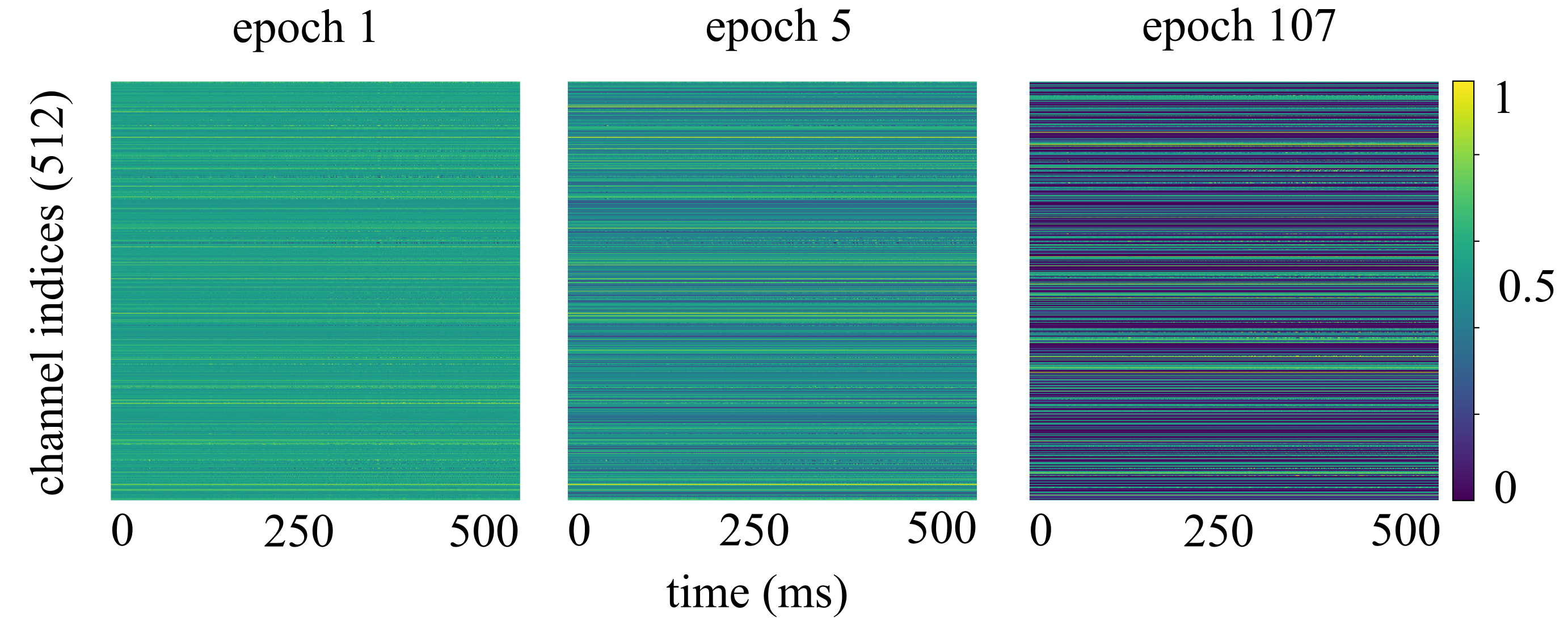}
 \vspace{-7mm}
    \caption{Evolution of the gate outputs for all the 512 channels from the first gated layer (for the same test utterance). }
    \label{gateoutputs}
 \vspace{-3mm}
\end{figure}

\begin{figure*}[!h]
\vspace{-7.5mm}
    \includegraphics[width=0.96\textwidth]{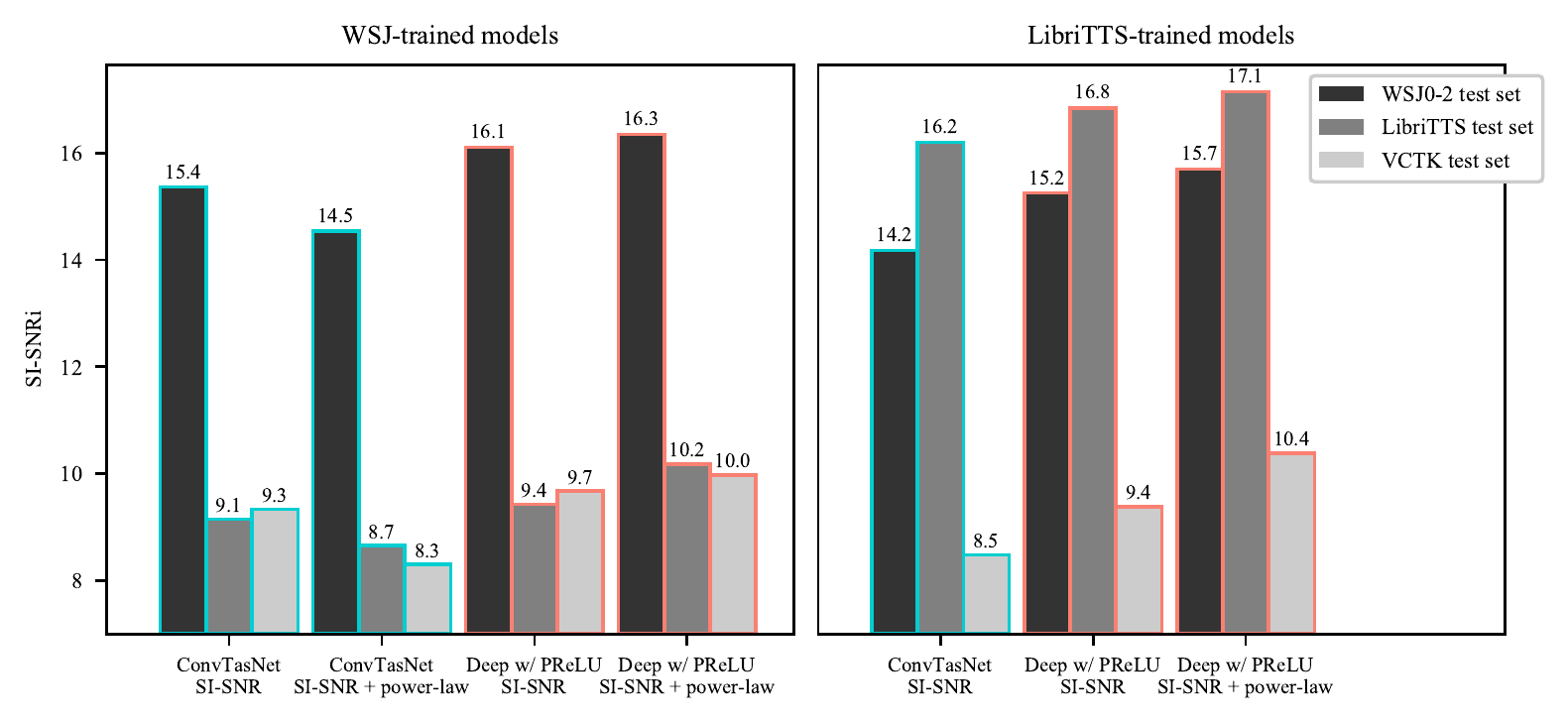}
    \vspace{-4mm}
    \caption{Cross-dataset evaluation on WSJ0-2mix, LibriTTS, and VCTK test sets.}
    \label{fig:crossdataset}
    \vspace{-4mm}
\end{figure*}

\subsection{Evaluation on the WSJ0-2mix Dataset}
In this section, we evaluate various models on the commonly used WSJ0 2-speaker (WSJ0-2mix) database \cite{merlscript}. Training (30 hours) and validation sets (10 hours) are created by randomly mixing utterances from 100 speakers 
at randomly selected SNRs between -5 and 5 dB. All waveforms are sampled at 8000 Hz. We report SI-SNR improvement (SI-SNRi) before and after speech separation on the test set (5 hours, 16 unseen speakers) in Table \ref{modelcomps}.

\begin{table}[!t]
\centering
 \caption{Summary of the studied variants and their performance on the WSJ0-2mix evaluation set. Objective function is SI-SNR.}
 \vspace{3mm}
\resizebox{\columnwidth}{!}{
\begin{tabular}{|c||c|c|c|c|c|}
\hline
 \multirow{2}{*}{\textbf{Model}} & \textbf{($I$)} & \textbf{Non-} & \textbf{\# first-layer} & \textbf{SI-SNRi} \\ 
   & \textbf{\# layers} & \textbf{linearity} & \textbf{filters} & \textbf{(dB)} \\ \hline  \hline

  Conv-TasNet & 1 & Linear & 512  & 15.4 \\  \hline

  Big-Conv-TasNet & 1 & Linear & 1024  & 15.3 \\  \hline 

   Deep w/ PReLU & 4 & PReLU & 512  & 16.1 \\  \hline

   Deep w/ dilation & 5 & PReLU & 512  & 16.0 \\  \hline

    Deep w/ GLU & 4 & GLU & 512  & \textbf{16.2} \\  \hline

\end{tabular}}
\vspace{-3mm}
    \label{modelcomps}
\end{table}

First, we successfully reproduced the original Conv-TasNet to build upon their result (row 1, Table \ref{modelcomps}). 
Second, note that the Big-Conv-TasNet \mbox{(row 2, Table \ref{modelcomps})}, a modified Conv-TasNet with double the number of kernel filters and four TCNs (instead of three) in the separator, is not able to outperform the original Conv-TasNet.
And third, the deep encoder/decoder \mbox{(row 3, Table \ref{modelcomps})} provides 0.7 dB improvement over the baseline. 
This result denotes the importance of the architecture itself because the objective metrics did not improve by simply increasing the capacity of the model.

With our first variant of the deep encoder/decoder, we investigated the potential impact of increasing the temporal context of the deep layers~---~via employing dilated convolutions.
Interestingly, the dilated convolutions in the deep encoder/decoder \mbox{(row 4, Table \ref{modelcomps})} did not further improve results, whereas \cite{luo2019conv} found dilations to be crucial for the separator. Considering this observation, along with the fact that the optimal encoder/decoder kernel length in Conv-TasNet is as short as 2 ms \cite{luo2019conv}, we hypothesize that the encoder tends to be optimized for modeling local patterns with high time resolution, while the separator might focus on learning longer temporal patterns. 

In our second variant of the deep encoder/decoder, we investigate the impact of further increasing its capacity with GLUs.
Our results 
\mbox{(row 5, Table \ref{modelcomps})} indicate that GLUs provide a very minor improvement over the basic deep encoder/decoder.
To further explore this result, we analyzed the gate outputs from the first gated layer (see Fig. \ref{gateoutputs}). 
Note that these gate outputs indicate the relative importance of each channel (upper branch in Figure \ref{fig:GLU}) in the layer.  In an early training stage, all gates are uniformly open. As training progresses, the model gradually ``closes" a few gates. After convergence, certain channels are clearly omitted~---~and, out of the remaining ones, some channels are much more dominant than others. This evolution was consistently observed throughout the test set, and shows that the gates learn to sift out less useful channels. 
We hypothesize that this behavior might be caused by the current learning strategy, forcing the model to start with a large set of candidates and gradually learning which channels work well and which channels can be omitted.  A more effective learning strategy might potentially lead to a reduced model size and shorter training~times.

\subsection{Cross-dataset Evaluation} \label{cdseval}
Here we explore the impact of training Conv-TasNet and the deep encoder/decoder on a larger, more diverse, training set: LibriTTS \cite{zen2019libritts}.  Our goal is to compare the SI-SNRi performance of these two architectures when using the WSJ and LibriTTS datasets for training~---~and the WSJ, LibriTTS, and VCTK \cite{vctk} datasets for evaluation. With this cross-dataset evaluation strategy, we aim to investigate the generalization capabilities of the studied models.

We assemble the LibriTTS and VCTK datasets by randomly mixing utterances at randomly selected SNRs between -5 and 5 dB, with waveforms sampled at 8000 Hz. These two additional datasets have the following characteristics:
\begin{itemize}\compresslist
\item For building the LibriTTS training set, we re-mix the train-clean-100 and train-clean-360 sets. Together, they comprise 245 hours of training data and 1151 speakers (553 female, 598 male).
Roughly speaking, it is about 8x the training data and 10x the number of speakers than WSJ0-2mix.  
The validation and evaluation sets are built by re-mixing dev-clean and test-clean, respectively.
The resulting validation set is 9 hours (40 speakers) and the evaluation set is 8.5 hours (39 speakers).
This dataset is used for training and evaluation purposes.
\item For VCTK, we randomly select 4000 pairs of utterances from the full dataset. This dataset is only used for evaluation purposes.
\end{itemize}

\noindent Fig. \ref{fig:crossdataset} shows the SI-SNRi evaluation results of several experiments. For each experiment, we evaluate on WSJ, LibriTTS, and VCTK.  In total, four of the experiments use WSJ for training, and three use LibriTTS. Among those, three are based on the original Conv-TasNet, and the rest employ the deep encoder/decoder.
Finally, we also experiment with two distinct objective functions: the standard SI-SNR objective, and an augmented SI-SNR + spectral power loss objective (defined in Eq. \ref{plaw}).
From Fig. \ref{fig:crossdataset}, some clear trends emerge:
\begin{enumerate} \compresslist
    \item Models trained on WSJ have a significant cross-dataset drop in performance when tested on LibriTTS (6.2 dB average) and VCTK (6.3 dB average), while models trained on LibriTTS have a relatively small drop in performance on WSJ ($1.7$ dB average) --- but still have a significant drop on VCTK ($7.3$ dB average).
    \item The deep encoder/decoder trained with SI-SNR provides a consistent performance improvement relative to the original Conv-TasNet --- of $0.65$ dB SI-SNRi on average for in-dataset evaluation; and $0.35$ dB and $0.95$~dB for WSJ-trained and LibriTTS-trained models, respectively, for cross-dataset evaluation.
    \item The power-law term consistently improves the performance of the deep encoder/decoder for both in-dataset ($0.25$ dB average) and cross-dataset ($0.65$ dB average) performance.  Compared to Conv-TasNet trained with SI-SNR, the average improvement is $1.2$ dB across all test results.  Interestingly, the power-law term has a negative impact on SI-SNRi for the original Conv-TasNet (-$0.9$ dB in-dataset, and -$0.7$ dB average cross-dataset). 
\end{enumerate}

\noindent 
Although the larger LibriTTS dataset improves cross-dataset performance on average, 
the results on VCTK suggest that increasing the scale and diversity of the training set alone may not be sufficient to ensure improved generalization.  Augmenting the deep encoder/decoder's objective with the power-law loss may have a regularizing effect, which could explain why it appears to help with generalization. 
We also noted some stability issues during training on very large datasets that were specific to the deep encoder/decoder when it is optimized solely with SI-SNR.  In all of our experiments, adding the power-law loss term (that seems to act as a regularizer) was sufficient to resolve these stability issues.
In line with this interpretation, it is possible that more tuning (e.g. reducing the $\beta$ parameter) may be required for Conv-TasNet since the model is smaller and might require less regularization.

\subsection{Deep Encoder/Decoder vs. Enhanced Separators}
Since other Conv-TasNet variants have focused on improving the separator~\cite{shi2019endfurca,shi2019deep,yang2019improved}, and we mostly focus on the deep encoder/decoder, we also compare the contribution of the proposed deep encoder/decoder with that of several enhanced separators in Table \ref{encvssep}. We compare our results for signal-to-distortion ratio improvement (SDRi) \cite{le2019sdr} against FurcaPy \cite{shi2019endfurca} and Yang et al.~\cite{yang2019improved}, both built on Conv-TasNet and with enhanced separators. FurcaPy \cite{shi2019endfurca} has recently achieved state-of-the-art results by employing parallel gated separators. On the other hand, Yang et al.~\cite{yang2019improved} is another recent work that enhances the separator by utilizing an embedding network and clustering. Their separator also takes advantage of STFT features, whereas our model learns only from waveforms. When comparing their results with ours, we note that by simply increasing the depth of the encoder/decoder one can achieve a similar improvement to Yang et al.~\cite{yang2019improved}. This result denotes the research value and potential of the encoder/decoder. However, we also note that by simply improving the encoder/decoder one cannot achieve state-of-the-art performance. These results seem to indicate that joint solutions based on improved encoder/decoders and improved separators are a promising next step to follow.

\begin{table}[]
\vspace{-2mm}
\caption{SDRi results on the WSJ0-2mix evaluation set for Conv-TasNet, the deep encoder/decoder, and enhanced separator models.}
 \vspace{3.5mm}

\centering
\begin{tabular}{|c||c|c|}
\hline
 \textbf{Model} & \textbf{SDRi (dB)} & \textbf{\# params}\\ \hline \hline
 Conv-TasNet & 15.6 & 5.0M \\ \hline
  Deep w/ PReLU & 16.6 & 9.7M \\ \hline
  Yang et al. \cite{yang2019improved} & 16.9 & 10M \\ \hline
  FurcaPy \cite{shi2019endfurca} & 18.4 & N/A \\ \hline
\end{tabular}
\vspace{-3mm}
\label{encvssep}
\end{table}
\section{Conclusion}
\label{sec:conclusion}
In this work, we present an empirical study where we evaluate the impact of several modifications to the original Conv-TasNet. 
We propose a deep (nonlinear) encoder/decoder variant that consistently outperforms the original (linear) encoder/decoder of Conv-TasNet. This result denotes the potential 
of improving the encoder/decoder, that is often overlooked.
We also investigate the relative impact of using a larger, more diverse training set --- in a cross-dataset evaluation setup designed to evaluate the generalization capabilities of the studied models.
While these experiments confirmed the consistency of the improvements provided by the deep encoder/decoder, these also highlighted the challenges of generalizing to unseen datasets.
We hope that this empirical study will lead to a deeper understanding of Conv-TasNet and inspire continued research into the generalization capabilities of end-to-end speech source separation models.

\section{Acknowledgments}
\label{sec:acknowledgments}
The authors thank Rich Graff and Nathan Swedlow for their help with the evaluation section and Roy Fejgin for his help with editing.
\newpage
\bibliographystyle{IEEEbib}
\bibliography{z_bibliography}
\end{document}